\documentclass[10pt,showpacs,preprintnumbers]{revtex4}
\usepackage{amsfonts}

\usepackage{color}
\usepackage{hyperref}

\def\com{\color{magenta}}

\newcommand{\oarX}[1]{\href{http://arxiv.org/abs/#1}{{\ttfamily\com #1}}}
\newcommand{\arX}[1]{\href{http://arxiv.org/abs/#1}{{\ttfamily\com arXiv:#1}}}

\def\barr{\begin{array}}
\def\earr{\end{array}}

\def\ben{\begin{equation}}
\def\een{\end{equation}}
\def\bs{\begin{subequations}}
\def\es{\end{subequations}}
\def\bena{\begin{eqnarray}}
\def\eena{\end{eqnarray}}

\def\SU{{\rm SU}}

\def\im{{\rm i}}

\newcommand{\eg}{\textit{e.g.}~}

\begin{document}
\title{Quantum cosmology from quantum gravity condensates: cosmological variables and lattice-refined dynamics}
\author{\bf Steffen Gielen}
\email{sgielen@perimeterinstitute.ca}
\affiliation{Perimeter Institute for Theoretical Physics, 31 Caroline Street North, Waterloo, Ontario N2L 2Y5, Canada} 
\author{\bf Daniele Oriti}
\email{doriti@aei.mpg.de}
\affiliation{Max Planck Institute for Gravitational Physics (Albert Einstein Institute), Am M\"uhlenberg 1, 14476 Golm, Germany, EU}

\begin{abstract}

In the context of group field theory condensate cosmology, we clarify the extraction of cosmological variables from the microscopic quantum gravity degrees of freedom. 
We show that an important implication of the second quantized formalism is the dependence of cosmological variables and equations on the {\em quantum gravitational atomic number $N$} (number of spin network vertices/elementary simplices). We clarify the relation of the effective cosmological equations with loop quantum cosmology, understood as an effective (hydrodynamic-like) approximation of a more fundamental quantum gravity theory. By doing so, we provide a fundamental basis to the idea of lattice refinement, showing the dependence of the effective cosmological connection on $N$, and hence indirectly on the scale factor. Our results open a new arena for exploring effective cosmological dynamics, as this depends crucially on the new observable $N$, which is entirely of quantum gravitational origin. \end{abstract}

\date{November 8, 2014}

\preprint{AEI-2014-032}

\maketitle

\section{Introduction}

One major challenge for background-independent approaches to quantum gravity has been the description of macroscopic, (approximately) continuous and almost spatially homogeneous universes like our own, and the derivation of manageable effective equations describing the dynamics of such universes within a given fundamental theory. Completing these steps is crucial for deriving predictions of such theories, to be compared with observations such as those of Planck and BICEP2 \cite{planck}. The challenge arises because background independence implies that the most natural `vacuum state' describes no geometry at all, and a macroscopic non-degenerate (metric) geometry is unlikely to be found as a perturbative excitation over this vacuum. One also generically has to turn discrete structures into approximately continuous ones within this background-independent context.

Recently \cite{gfcpapers}, a major step towards addressing this challenge was completed within the group field theory (GFT) approach to quantum gravity \cite{GFT}, itself a second quantized formulation of loop quantum gravity (LQG) \cite{LQG, daniele2q}, with spin network vertices or elementary simplices as \lq quanta\rq\, of the GFT field, or \lq atoms of quantum space\rq. It was shown that many-atom {\em condensate} states in GFT, akin to coherent or squeezed states used in Bose--Einstein condensates, describe macroscopic spatially homogeneous universes. The effective equations defining the dynamics of these states, extracted directly from the fundamental GFT quantum dynamics, can be interpreted in terms of (nonlinear) quantum cosmology equations on minisuperspace. This result was shown to be structural and very general. In Ref.~\cite{gfcpapers} an example was given in which a certain choice of condensate state, with some assumptions on the GFT action, gives a linear effective equation whose semiclassical (WKB) limit is, in the isotropic case, the Friedmann equation for homogeneous isotropic general relativity. This result was obtained both for Riemannian and Lorentzian metric signature, both in vacuum and with a massless scalar field (see the second paper in Ref.~\cite{gfcpapers} for a detailed discussion of all these cases). In this scheme, spacetime emerges from the collective behaviour of a quantum condensate of pre-geometric, non-spatio-temporal degrees of freedom. Therefore, it represents an explicit, even if still tentative and partial, realisation of several suggestions coming from analogue gravity models in condensed matter theory \cite{analog}.

The purpose of this paper is to clarify further, in this context of GFT condensate cosmology, the extraction of cosmological variables from the microscopic degrees of freedom, which is the crucial step in interpreting the effective quantum cosmological equations. We use the {\em second quantization} formalism of GFT, namely the treatment of LQG spin network vertices as {\em indistinguishable} bosonic `quantum gravity atoms', and show that an important implication of a second quantized formalism is the dependence of cosmological variables and equations on the {\em atomic number $N$}. Using this observation, we explain the limitations of the WKB approximation used in Ref.~\cite{gfcpapers}. We can then derive the precise relation of our effective cosmological equations to the dynamics of loop quantum cosmology (LQC) \cite{LQC}: the dependence of the cosmological spin connection on $N$, and hence indirectly also on observables such as the scale factor, gives a fundamental basis to the idea of lattice refinement \cite{refine} and \lq improved dynamics\rq\, \cite{improv} in LQC. Our results open a new arena for exploring effective cosmological dynamics, as the new observable $N$ is entirely of quantum gravitational origin. It can explain and affect several elements of the effective cosmological dynamics, being thus an important ingredient of model building and analysis in (quantum) cosmology, understood as an effective (hydrodynamic-like) approximation of a more fundamental quantum gravity theory. A further result is that cosmological effective equations can be obtained from the fundamental quantum gravity dynamics in terms of expectation values of collective observables, not relying on any semiclassical approximation.

We find that continuum cosmological quantities can be associated either to {\em total} (extensive) or to {\em averaged} (intensive) observables of many-atom states, and argue that out of the canonically conjugate variables corresponding to a flux and a connection, the first must be `total' while the second is `averaged'. It follows that the relation between a macroscopic gravitational connection and the GFT group variables (representing parallel transports of a connection) must involve the average atomic number $\langle\hat{N}\rangle$. Indeed, the variable appearing in the effective cosmological equations can be identified with $\sin(\mu\,\omega)$ where $\mu\propto \langle\hat{N}\rangle^{-1/3}$. A change of the atomic number under time evolution then realizes a dynamical lattice refinement: the emergence of new quantum geometric degrees of freedom affects the effective cosmological dynamics.

The physical interpretation of this lattice refinement depends on how the quantum gravitational atomic number $N$, which has no analogue in the classical continuum theory, is related to cosmological quantities such as the scale factor. This relation is encoded in the cosmological `wavefunction' (a hydrodynamic variable in quantum gravity), and is itself {\em dynamical}. We look at two cases. In the first, the atomic number is (approximately) fixed and $\mu$ becomes a (constant) parameter. This scenario, implicitly assumed in Ref.~\cite{gfcpapers}, reproduces the constructions of `old' LQC \cite{lqcold}. In the second, we assume that the average volume per atom of geometry remains constant under time evolution: the average atomic number scales with the total volume, $\mu\propto\frac{1}{a}$ and the holonomy-corrected term replacing the connection $\omega$ is $\sin(\frac{a_0}{a}\,\omega)$. This reproduces precisely the form of holonomy corrections in the `improved dynamics' prescription in LQC \cite{improv}. The two scenarios should be considered as special cases of more general functional relations $N(a)$. In GFT condensate cosmology, this relation is ultimately a computable {\em result} of the fundamental dynamics of the theory and no additional assumptions on its form are needed. While this explicit computation is left to future work, the purpose of our paper is to show, by two explicit examples, how the form of $N(a)$ affects directly the effective cosmological dynamics and the resulting possible cosmological scenarios that can then be confronted with observation.

\section{Cosmological observables for GFT condensates} 

In usual non-relativistic many-particle quantum physics, the canonically conjugate single-particle operators $\hat{x}_i$ (position) and $\hat{p}_j$ (momentum) extend to a `total position' operator $\hat{X}_i$ and a total momentum operator $\hat{P}_i$ on the Fock space, which satisfy
\ben
[\hat{X}_i,\hat{P}_j] = \im\hbar\,\delta_{ij}\,\hat{N}\,,
\een
where $\hat{N}$ is the number operator. These operators are {\em not} canonically conjugate and `total position' has a rather unclear physical meaning, in contrast to the total momentum. The two issues are related: out of two canonically conjugate quantities, one is typically extensive and one is intensive in the particle number. One-body operators in second quantization, instead, are always extensive.

At fixed particle number $N$, $\frac{1}{N}\hat{X}_i$ defines a centre-of-mass position operator. However, a Fock space operator $\widehat{N^{-1}}$ is not naturally defined, as $\hat{N}$ contains zero in its spectrum. One can instead define the intensive quantity `average centre-of-mass position' as an expectation value, $x^{{\rm c.o.m.}}_i = \langle\hat{X}_i\rangle/\langle\hat{N}\rangle$.

The same discussion goes through for quantum gravity in the GFT context (see also the analysis in Ref.~\cite{coherent}). In the formulation of classical general relativity in terms of Ashtekar--Barbero variables \cite{asht}, the canonically conjugate continuum variables are the gravitational ${\rm SU}(2)$ connection $A_i^a$ and the `inverse triad' $E^j_b$. The inverse triad specifies a Riemannian 3-geometry on a spatial hypersurface: it is a density-weighted orthonormal frame which defines the (inverse) 3-metric by $E^i_a E^j_b \delta^{ab}=:(\det g)g^{ij}$. The Ashtekar--Barbero connection $A$ is then given by $A_i^a=\Gamma_i^a[E]+\gamma\, K^i_a$ in terms of the unique torsion-free Levi--Civita connection $\Gamma_i^a[E]$ associated to $E$ and the extrinsic curvature $K^i_a$; $\gamma\neq 0$ is a (real) free parameter called the Barbero--Immirzi parameter. In the construction of LQG \cite{LQG}, these continuum fields are discretized by integration over links and surfaces, giving for each link a group element $g$ (the parallel transport of $A$) and a Lie algebra element $B^i$ (the flux of $E$ through a dual surface), with {\em holonomy-flux algebra}
\ben
\{g,B^i\}=-(8\pi \gamma {\rm G_N})\tau^ig\,,\;\{B^i,B^j\}=-(8\pi \gamma {\rm G_N})\,{\epsilon^{ij}}_kB^k\,.
\een
Here $\tau^i$ are a basis of the $\SU(2)$ Lie algebra, {\em e.g.} $\tau^i=\frac{{\rm i}}{2}\sigma^i$, and ${\rm G_N}$ is Newton's constant. In the Fock space picture of 4$d$ GFT, four copies of $g$ and $B^i$ become the basic phase space variables parametrizing single-atom states (of individual building blocks of quantum space); for each copy, the corresponding single-atom operators satisfy 
\ben
[\hat{g},\hat{B}^i]=-\im\kappa\tau^i\hat{g}\,,\;[\hat{B}^i,\hat{B}^j]=-\im\kappa\,{\epsilon^{ij}}_k\hat{B}^k\,,
\label{algebra}
\een
where $\kappa:=8\pi\gamma \hbar{\rm G_N}$ has dimensions of area. 

The GFT Fock space is constructed from a vacuum state $|\emptyset\rangle$ describing a completely degenerate geometry, analogous to the standard LQG vacuum \cite{LQG}. While $\hat\varphi(g_I)|\emptyset\rangle=0$, the conjugate operator $\hat\varphi^{\dagger}(g_I)$ creates an `atom of space' labelled by group elements $g_I$ ($I=1,\ldots,4$). With bosonic statistics for $\hat\varphi$, many-atom states can be constructed by repeated actions of $\hat\varphi^{\dagger}(g_I)$ on $|\emptyset\rangle$. Such states correspond to spin networks, with vertices as basic quanta, and can be interpreted as triangulations labelled by the same type of algebraic data. Their interpretation in terms of a continuum metric may require an embedding into a manifold and it is not guaranteed in general. See Refs.~\cite{gfcpapers,GFT,LQG} for details of the GFT states, their geometric interpretation, the relation to LQG, and the GFT dynamics.

On the GFT Fock space, the operators ($\hat{g},\hat{B}^i$) extend to a `total group element', defined in terms of a coordinate system on $\SU(2)$, and a total flux $\hat{b}^i$. We choose coordinates $\vec\pi$ by
\ben
g =  \sqrt{1-\vec{\pi}[g]^2}\,{\bf 1} - \im\vec{\sigma}\cdot\vec\pi[g]\,,\quad|\vec{\pi}[g]|\le 1\,.
\een
The `total group coordinate' operators (analogous to the `total position' above)
\ben
\hat{\Pi}[g_I] = \int ({\rm d}g)^4\; \vec\pi[g_I]\;\hat{\varphi}^{\dagger}(g_J)\hat{\varphi}(g_J)
\label{groupel}
\een
and total flux operators, represented as right-invariant vector fields on $\SU(2)$
\ben
\hat{b}^i_I = \im\kappa\int ({\rm d}g)^4\; \hat{\varphi}^{\dagger}(g_J)\frac{{\rm d}}{{\rm d}t}\hat{\varphi}\left(\exp\left(\tau^i_I\,t\right)g_J\right)\Big|_{t=0}\,,
\label{fluxop}
\een
are well-defined on the Fock space. 
This total flux is non-commutative; we interpret its commutative limit
\ben
\hat{f}^i_I = \im\kappa\int ({\rm d}g)^4\; \hat{\varphi}^{\dagger}(\pi[g_J])\frac{\partial}{\partial\pi_i^I}\hat{\varphi}(\pi[g_J])
\label{fluxdef}
\een
to be the macroscopic flux variable of direct cosmological interpretation.

As in the previous example, the flux defines a naturally extensive quantity, while the `total group element' carries no obvious interpretation. We can however define `average group coordinate' operators by
\ben
\hat\Pi[g_I]^{{\rm av.}} = \hat{\Pi}[g_I]/ \langle \hat{N} \rangle\,.
\een
The eigenvalues of $\hat\Pi_I^{{\rm av.}}\equiv\hat\Pi[g_I]^{{\rm av.}}$ satisfy $|\langle\hat\Pi_I^{{\rm av.}}\rangle|\le 1$.

The {\em total} fluxes $\hat{b}_I$ and the {\em averaged} group coordinates $\hat\Pi_I^{{\rm av.}}=\frac{1}{\langle\hat{N}\rangle}\hat{\Pi}_I$, analogous to total momentum and centre-of-mass position, characterize the condensate. In particular, only {\em averaged holonomies} can be interpreted consistently as macroscopic holonomies: noting that the parallel transport over a path of coordinate length $\mu$ in $j$-direction, with approximately constant connection, is
\ben
\mathcal{P}\exp\int\omega \approx  \cos(\mu|\omega_j|)\,{\bf 1} + \frac{\omega_j}{|\omega_j|}\sin(\mu|\omega_j|)
\een
with $\omega_j\in\mathfrak{su}(2)$,  the averaged group coordinates can thus be interpreted as the parallel transport of a connection
\ben
\omega=\im\vec\sigma\cdot\vec\omega\,,\quad\mu\,\vec\omega:=-\frac{\langle\vec{\Pi}\rangle}{|\langle\vec{\Pi}\rangle|}\,\arcsin\frac{|\langle\vec{\Pi}\rangle|}{N}\,.
\label{ident}
\een
Eq.~(\ref{ident}) is a change of variables $(\vec\Pi,N)\rightarrow(\vec\omega,N)$.

Fixing $\mu$ defines a coordinate system in which $\omega$ is given. In Ref.~\cite{gfcpapers}, $\mu$ was taken to be the coordinate length of a `fundamental link' associated to an elementary quantum of geometry, and taken as constant. However, as $N$ appears explicitly in Eq.~(\ref{ident}), it appears unnatural to assume that $\mu$ should be independent of $N$. 
A more natural coordinate system is one in which the whole condensate is extended over a region of fixed coordinate length. Each quantum then occupies an average coordinate volume $\propto 1/N$, and the coordinate length for a quantum is $\mu\propto N^{-1/3}$. Adopting this coordinate system (in itself of no physical content) is convenient for linking the effective cosmological equations arising from GFT condensates and the formalism of loop quantum cosmology. The so-defined collective variables correspond to the macroscopic, cosmological variables for the GFT condensate.

\section{Interpretation of effective cosmological equations}

In Ref.~\cite{gfcpapers} it was shown that the dynamics of condensate states in GFT can be reduced, within certain approximations, to effective quantum cosmology equations. These arise from Schwinger--Dyson equations of the GFT, which take the general form
\ben
\left\langle \frac{\delta\mathcal{O}[\varphi,\bar\varphi]}{\delta\bar\varphi(g_I)}-\mathcal{O}[\varphi,\bar\varphi]\frac{\delta S[\varphi,\bar\varphi]}{\delta\bar\varphi(g_I)}\right\rangle = 0
\label{sdys}
\een
for any functional $\mathcal{O}$ of the GFT field $\varphi$ and its complex conjugate, with fundamental dynamics defined by an action $S$ (see \cite{GFT,LQG} for the most developed and promising models). Eq.~(\ref{sdys}) holds in the vacuum state, for all $\mathcal{O}$. Requiring Eq.~(\ref{sdys}) for certain $\mathcal{O}$ encodes the requirement for a GFT condensate state to give a good approximation to a non-perturbative vacuum (see Ref.~\cite{lorenzo} for further analysis of this approximation). The key result of Ref.~\cite{gfcpapers}, at the dynamical level, was that Eq.~(\ref{sdys}), for simple choices of $\mathcal{O}$ and for an approximate vacuum state given by a GFT condensate state such as
\ben
|\sigma\rangle \propto\exp\left(\hat\sigma\right)|0\rangle\,,\quad\hat\sigma := \int ({\rm d}g)^4\, \sigma(g_I)\hat\varphi^{\dagger}(g_I) \,,\label{cond}
\een gives a quantum cosmology-like equation for the cosmological `wavefunction' $\sigma$ (similar to those obtained in Ref.~\cite{nonlinLQC}). Here we want to interpret Eq.~(\ref{sdys}) directly in terms of expectation values of second quantized operators corresponding to the kinetic and interaction terms of the fundamental GFT action, using again condensate states with a cosmological interpretation.

For $\mathcal{O}=\bar\varphi(g_I)$, Eq.~(\ref{sdys}) becomes, integrating over $g_I$,
\ben
\left\langle \int ({\rm d}g)^4 \; \bar\varphi(g_I)\frac{\delta S[\varphi,\bar\varphi]}{\delta\bar\varphi(g_I)}\right\rangle = \left\langle \int ({\rm d}g)^4 \; \frac{\delta\bar\varphi(g_I)}{\delta\bar\varphi(g_I)}\right\rangle\,.
\een
Passing to the operator formalism and choosing normal ordering, the delta distribution $\delta\bar\varphi/\delta\bar\varphi$ disappears and
\ben
\left\langle \int ({\rm d}g)^4 \; \hat\varphi^{\dagger}(g_I)\frac{\delta \hat{S}[\hat\varphi,\hat\varphi^{\dagger}]}{\delta\hat\varphi^{\dagger}(g_I)}\right\rangle = 0 \, .
\label{expval}
\een
Eq.~(\ref{expval}) has to hold for any state that defines a GFT vacuum. Writing the action as $\hat{S}= \hat{K}+\hat{\mathcal{V}}$ for a second quantized kinetic operator
\ben
\hat{K}[\hat\varphi,\hat\varphi^{\dagger}]=\int ({\rm d}g)^4({\rm d}g')^4\;\hat{\varphi}^\dagger(g_I)\mathcal{K}(g_I,g'_I)\hat{\varphi}(g'_I)
\een
and interactions $\hat{\mathcal{V}}$, Eq.~(\ref{expval}) becomes $\langle\hat{K}\rangle+\ldots = 0$ with the second term linear in $\hat{\mathcal{V}}$. We are now interested in the situation in which this second term can be set to zero. This can be an exact result for certain states \cite{gfcpapers} or, more generally, one could consider a weak-coupling limit neglecting the interactions. We then require a GFT condensate, {\em e.g.} (\ref{cond}), to satisfy $\langle\hat{K}\rangle=0$.

To make the corresponding cosmological dynamics explicit, we need to make a choice for $\mathcal{K}$. As in Ref.~\cite{gfcpapers} (and as motivated by GFT renormalisation \cite{GFTrenorm}), we choose $\mathcal{K}=\sum_I \Delta_{g_I} + m^2$ where $\Delta_{g}$ is the Laplace--Beltrami operator on $\SU(2)$ and $m^2$ is a coupling constant. $\Delta_g$ can be expressed as a combination of right-invariant vector fields and hence of non-commutative fluxes as in Eq.~(\ref{fluxop}). For easier comparison with continuum classical gravity, we instead express $\Delta_g$ in terms of partial derivatives, {\em i.e.} of the commutative total fluxes of Eq.~(\ref{fluxdef}). We have
\ben
\Delta_{g_I}=(\delta_{ij}-\pi_i^I\pi_j^I)\frac{\partial}{\partial\pi_i^I}\frac{\partial}{\partial\pi_j^I}-3\pi_i^I\frac{\partial}{\partial\pi_i^I}\,.
\een
The equation $\langle \hat{K}\rangle = 0$ can be rewritten as
\ben
\sum_I\left\langle \hat{f_I\cdot f_I}-\hat{(\Pi_I\cdot f_I)^2}-3\im\kappa(\hat{\Pi_I\cdot f_I})\right\rangle-m^2\kappa^2\langle\hat{N}\rangle = 0 
\label{expval3}
\een
where we have defined the one-body operators on the GFT Fock space
\bena
\hat{f_I\cdot f_I} &=& -\kappa^2\int ({\rm d}g)^4\; \hat{\varphi}^{\dagger}(\pi[g_I])\delta_{ij}\frac{\partial}{\partial\pi_i^I}\frac{\partial}{\partial\pi_j^I}\hat{\varphi}(\pi[g_I])\,,
\label{fsqop}
\\\hat{(\Pi_I\cdot f_I)^2} &=& -\kappa^2\int ({\rm d}g)^4\; \hat{\varphi}^{\dagger}(\pi[g_I])\pi_i^I\pi_j^I\frac{\partial}{\partial\pi_i^I}\frac{\partial}{\partial\pi_j^I}\hat{\varphi}(\pi[g_I])\,,
\\\hat{\Pi_I\cdot f_I} &=& \im\kappa\int ({\rm d}g)^4\; \hat{\varphi}^{\dagger}(\pi[g_I])\pi^I_i\frac{\partial}{\partial\pi_i^I}\hat{\varphi}(\pi[g_I])\,.
\label{pfop}
\eena
In a condensate (such as (\ref{cond})), all quantum gravitational atoms are in the same configuration and 
\ben
\langle \hat{f_I\cdot f_I} \rangle \approx \frac{1}{\langle\hat{N}\rangle} \langle \hat{f}_I \rangle \cdot \langle \hat{f}_I \rangle\,,\quad{\rm etc.}\,,
\label{approx}
\een
so that Eq.~(\ref{expval3}) becomes approximately
\ben
\sum_I \langle \hat{f}_I \rangle \cdot \langle \hat{f}_I \rangle-\frac{(\langle \hat{\Pi}_I \rangle \cdot \langle \hat{f}_I \rangle)^2}{\langle \hat{N}\rangle^2}-3\im\kappa\langle\hat\Pi_I\rangle\cdot \langle\hat{f}_I\rangle\approx m^2\kappa^2\langle\hat{N}\rangle^2\,.
\een

We can now identify expectation values with the variables of homogeneous, isotropic GR. While by using condensate states such as (\ref{cond}) we are already assuming homogeneity, we impose the property of isotropy on expectation values: we require that in an identification $\langle \hat{f}_I \rangle = T_I\,a^2$, where $a$ is the scale factor and $T_I\in\mathfrak{su}(2)$ with $T_I\cdot T_I = O(1)$, $T_I$ are state-dependent constants (encoding information about the shape of the elementary building blocks) while $a$ can be time-dependent, so that the resulting effective dynamics can be intepreted as a Friedmann equation for the variable $a$. Similarly, for the effective connection we have $\langle \hat{\Pi}_I \rangle/\langle\hat{N}\rangle = V_I  \sin(\mu\,\omega)$ for $V_I\in\mathfrak{su}(2)$ again with $V_I\cdot V_I = O(1)$, in terms of a single spin connection variable $\omega$. We obtain 
\ben
\frac{k-\sin^2(\mu\omega)}{a^2}-\frac{3\im\kappa}{a^4}\alpha N\sin(\mu\omega)-\frac{m^2\kappa^2N^2}{\beta a^6} \approx 0\,,
\label{fried}
\een
with $N=\langle\hat{N}\rangle$, and $k,\alpha,\beta$ are shorthands for combinations of $T_I\cdot T_I$, $T_I\cdot V_I$, etc. Viewed as a cosmological Friedmann equation, we find the usual gravitational term, including the same holonomy corrections as in LQC, and two terms depending on $N$ whose interpretation will depend on the exact relation between $N$ and the cosmological observables. Note the appearance of an apparently imaginary term which is due to the introduction of the operators (\ref{fsqop})-(\ref{pfop}) which, although having a clear cosmological interpretation, are not Hermitian on the GFT Fock space (their combination given by the Laplacian however is). One possible way of interpreting the appearance of an imaginary term in Eq.~(\ref{fried}) is as related to a Wick rotation: in Ref.~\cite{gfcpapers}, the analogue of Eq.~(\ref{fried}) was interpreted as describing Riemannian signature geometries. The passage to Lorentzian signature, at least when $\mu\omega\ll 1$, corresponds to $\omega\mapsto\pm\im\,\omega$ making all terms in Eq.~(\ref{fried}) real. Complex metric variables of course have a long history in quantum cosmology, \eg in semiclassical approximations \cite{hartlhawk}.

We emphasise that this dependence of the effective cosmological dynamics on $N$ is a new genuine quantum gravity effect, and that all the corrections to the Friedmann equation above are {\em derived} from the fundamental GFT dynamics. Indeed, our main observation is that effective cosmological equations for quantum gravity condensates necessarily depend on the atomic number $N$. In Ref.~\cite{gfcpapers}, an equation including only the first term of Eq.~(\ref{fried}) arose from a WKB approximation for the wavefunction $\sigma$; the WKB expansion in derivatives appeared to be an expansion in $\kappa/{a^2}$, indeed tiny for macroscopic $a$. 

The physical viability of the WKB approximation could be questioned as it assumes that the individual quanta of space are themselves semiclassical. One can study the deviations of explicit solutions from WKB behaviour \cite{steffegfc}. The WKB limit also neglected any dependence on $N$, simply identifying the total (extensive) operators with cosmological observables. We then understand its failure from a different angle here: when including the scaling with $N$, Eq.~(\ref{fried}) appears to be an expansion in powers of $\kappa N/{a^2}$, the inverse {\em average} area in (Planck) units set by $\kappa$. This need not be small even in the semiclassical case. 

Eq.~(\ref{fried}) arises from taking an expectation value in the condensate state. Using Eq.~(\ref{fried}) to describe the cosmological dynamics of the condensate assumes that relative fluctuations remain small, so that one can focus on expectation values. To study this property for specific states is a subtle issue, even in the context of LQC \cite{reply}, but it is an additional condition to be imposed on GFT condensates.

In order to connect effective equations like Eq.~(\ref{fried}) to classical general relativity or to LQC, we need to relate the new QG observable $N$ to geometric observables, such as the scale factor. This relation is encoded in the condensate wavefunction, and can be {\em computed} for any given solution of the effective dynamics. Here, we consider two interesting possible regimes. The first is when the condensate has an approximately constant atomic number $N$, that can thus be treated as an additional parameter. One could fix $N$ to be exactly constant by working in the canonical ensemble. We recover the variables of the `old' version of LQC \cite{lqcold}: the holonomy-corrected expression replacing $\omega$ is simply $\sin(\mu_0 \omega)$ for constant $\mu_0$.
Then the two terms in Eq.~(\ref{fried}) describe a ``radiation'' term that also depends on the connection, and a stiff matter term. 

A second regime is when the expansion of the Universe proceeds by an increase in the number of QG atoms, with constant average volume per atom. One can restrict to this regime as well by an appropriate choice of GFT ensemble of states. Then one has $N\propto a^3$ and $\mu=\frac{a_0}{a}$ for some $a_0$, so that the combination of $a$ and $\omega$ in the effective Friedmann equation is now $\sin(\frac{a_0}{a}\omega)$. This is precisely as in the `improved dynamics' prescription of LQC \cite{LQC, improv}, which is here {\em derived} from the fundamental dynamics of condensates in a second quantized version of LQG, given by GFT. For $N=(a/a_0)^3$, Eq.~(\ref{fried}) becomes
\ben
\frac{k-\sin^2(\frac{a_0}{a}\omega)}{a^2}-\frac{3\im\kappa}{a\,a_0^3}\alpha \sin\left(\frac{a_0}{a}\omega\right)-\frac{m^2\kappa^2}{\beta a_0^6} \approx 0
\label{fried2}
\een
or for $\frac{a_0}{a}\omega\ll 1$, 
\ben
\frac{k}{a_0^2}-\frac{\omega^2}{a^2}-\frac{3\im\kappa\alpha}{a_0^4}\omega-\frac{m^2\kappa^2}{\beta a_0^6}\frac{a^2}{a_0^2} \approx 0\,.
\label{fried3}
\een

In this regime, the $k$-dependent term appears as an effective cosmological constant. The term linear in the connection and the infrared modification growing as $a^2$ do not have a clear physical interpretation and there would be very strong observational constraints in particular on the last term which dominates at late times, basically forcing the `mass' coupling $m^2$ in the GFT kinetic term to vanish. The purpose of this paper is to explore different possibilities for effective Friedmann equations assuming different scenarios for the relation between $N$ and the scale factor $a$. More work is required to justify these from the fundamental theory, but once this has been achieved, the cosmological predictions coming from equations like (\ref{fried3}) can be confronted with observation. This comparison with observation will then be very useful for putting constraints on the possible choices of the fundamental GFT dynamics.

\section{Conclusion}

To obtain the effective cosmological dynamics (\ref{fried}), we have made a choice for the kinetic operator $\mathcal{K}$, but the general argument extends to any $\mathcal{K}$ that has second derivatives in the group variables. The precise form of Eq.~(\ref{fried}) would be however different for a different choice of $\mathcal{K}$, leading to a different cosmological interpretation of the corresponding terms. More corrections would come from the GFT interactions, which we have neglected here. 

Beside the specific form obtained, the main result is that the identification of the effective dynamics depends rather crucially on the behaviour of the atomic number $N$. This behaviour depends on the choice of ensemble and on the dynamical properties of the underlying field theory; moreover, it may change under a phase transition \cite{GFTfluid}, and different phases may be described by very different effective dynamics. The very presence of the quantity $N$ in the effective dynamics is a totally general and new feature, independent of the details of the model.

In particular, the relation $\mu\propto N^{-1/3}$ results from a kinematical identification of the condensate observables, and not from any specific choice of GFT dynamics. 
All aspects of the dynamics of LQC, including the precise form of holonomy corrections, can now be derived from microscopic quantum gravity (see also work aimed at similar goals within canonical LQG \cite{alescicianfrani} and spin foams \cite{SFcosmology}), due to the appearance of the new quantum gravity observable $N$ in effective cosmological equations. The mechanism itself gives a microscopic dynamical origin of lattice refinement \cite{refine} in LQC, as it suggests a dynamical change in the number of degrees of freedom of quantum geometry with the evolution of the Universe.
Further results strengthening the link between GFT condensate cosmology and lattice refined LQC are given in Ref.~\cite{gianluca}.

This gives further weight to the program of Ref.~\cite{gfcpapers} for deriving quantum cosmology equations from the dynamics of GFT condensate states, thus from a many-body quantum system of fundamental QG degrees of freedom.  

\section*{Acknowledgements}

We thank L. Sindoni and G. Calcagni for discussions. Research at Perimeter Institute is supported by the Government of Canada through Industry Canada and by the Province of Ontario through the Ministry of Research and Innovation. D.O. acknowledges financial support from the A. von Humboldt Stiftung with a Sofja Kovalevskaja Award.

\end{document}